\documentclass[letterpaper, 10 pt, conference]{ieeeconf}  

\IEEEoverridecommandlockouts                              

\usepackage{graphicx} 
\usepackage{amsmath} 
\usepackage{amssymb}  
\usepackage{float}
\usepackage{booktabs}
\usepackage{tensor}
\usepackage{accents}
\usepackage{mathtools}
\usepackage{algorithm}
\usepackage{algorithmic}
{
    \newtheorem{assum}{Assumption}
    \newtheorem{thrm}{Theorem}
    \newtheorem{lemma}{Lemma}
}
\usepackage{wasysym}

\newcommand{\ignore}[1]{}

\newcommand{\bma}[1]{\left[\begin{array}{#1}}
\newcommand{\ema}{\end{array}\right]}

\DeclareMathAlphabet{\mbf}{OT1}{ptm}{b}{n}

\def\fdotb{{\raisebox{-0.6ex}{ \kern0.2ex\raisebox{0.8ex}{\tiny $\hspace*{-1ex}\circ$}}}}
\def\fddotb{{\raisebox{-0.6ex}{ \kern0.2ex\raisebox{0.8ex}{\tiny $\hspace*{-1ex}\circ\circ$}}}}

\newcommand{\utimes}{ {\raisebox{-0.6ex}{ \kern-1.0ex\raisebox{0.6ex}{ \small $\mathsf{v}$}}} } %
\newcommand{\beq}{\begin{equation}}
\newcommand{\eeq}{\end{equation}}
\newcommand{\bdis}{\begin{displaymath}}
\newcommand{\edis}{\end{displaymath}}
\newcommand{\beqarray}{\begin{eqnarray}}
\newcommand{\eeqarray}{\end{eqnarray}}
\newcommand{\beqarraynn}{\begin{eqnarray*}}
\newcommand{\eeqarraynn}{\end{eqnarray*}}

\title{\LARGE \bf Smooth Logic Constraints in Nonlinear Optimization \\ and Optimal Control Problems}

\author{Jad Wehbeh$^{1}$ and Eric C. Kerrigan$^{2}$
\thanks{$^{1}$Jad Wehbeh is with the Department of Electrical and Electronic Engineering, Imperial College London, SW7 2AZ, UK
        {\tt\small j.wehbeh22@imperial.ac.uk}}%
\thanks{$^{2}$Eric C. Kerrigan is with the Department of Electrical and Electronic Engineering and the Department of Aeronautics, Imperial College London,
        SW7 2AZ, UK
        {\tt\small e.kerrigan@imperial.ac.uk}}
}

\begin{document}

\maketitle
\thispagestyle{empty}
\pagestyle{empty}

\setlength{\textfloatsep}{5pt}
\setlength{\intextsep}{5pt}
\setlength{\dbltextfloatsep}{5pt}

\begin{abstract}
    In some optimal control problems, complex relationships between states and inputs cannot be easily represented using continuous constraints, necessitating the use of discrete logic instead. This paper presents a method for incorporating such logic constraints directly within continuous optimization frameworks, eliminating the need for binary variables or specialized solvers. Our approach reformulates arbitrary logic constraints under minimal assumptions as max-min constraints, which are then converted into equivalent smooth constraints by introducing auxiliary variables into the optimization problem. We demonstrate the effectiveness of this method on two planar quadrotor control tasks with complex logic constraints. Compared to existing techniques for encoding logic in continuous optimization, our approach achieves faster computational performance and improved convergence to feasible solutions.
\end{abstract}


\section{Introduction}
\subsection{Background and Motivation}

In order to address the complexity of modern engineering challenges, optimal control methods have emerged as a popular and effective option for balancing performance and cost subject to user-defined and system-dependent constraints. These approaches generate control trajectories by solving optimization problems, which are usually formulated with smooth dynamics and constraints \cite[Chap.~1]{vinter2010optimal}, \cite[Chap.~12]{nocedal2006numerical}. However, many real-world applications involve complex relationships between the system variables that are not easily captured by smooth or continuous constraints and require the explicit modeling of logical links. This includes modeling contact in legged robotics \cite{wensing2023optimization}, energy price changes in power markets \cite{carli2022robust}, and thruster limitations in spacecraft \cite{malyuta2023fast}. These applications necessitate introducing binary variables to describe the logical constraints, resulting in mixed-integer problems with both continuous and discrete variables.

Unfortunately, using a mixed-integer solver in an optimal control context can be undesirable or inconvenient, particularly when the problem is nonlinear or the number of integer variables is small compared to the continuous dynamics. Standard nonlinear optimization solvers are generally more accessible than mixed-integer approaches and scale better to larger problem sizes than such techniques~\cite{kronqvist2019review, liberti2019undecidability}. Furthermore, mixed-integer methods tend to be significantly slower than continuous differentiable reformulations when such reformulations are feasible~\cite{kirches2020approximation}. Mixed-integer solvers typically provide the advantage of computing global solutions for the integer component of the optimization problem \cite{burer2012non}, but this comes at a significant computational cost, which is often unnecessary in the context of optimal control, where local solutions are usually sufficient. This is particularly true when the continuous part of the control problem is already non-convex, making the computation of global solutions for the entire problem highly impractical. Thus, there is a clear need for alternative methods to express logical relations in optimization problems without relying on binary variables or mixed-integer techniques.

\subsection{State of the Art}

Several approaches have been proposed for the modeling of logic constraints within the context of optimal control. The simplest formulations describe the constraints directly using integer variables \cite{bertsimas2021unified}, or use the big-M approach \cite[Chap.~4]{bazaraa2011linear} or complementarity constraints to capture the logical relations in the problem \cite{belotti2016handling}. However, all of these methods introduce discrete optimization variables, which necessitate the use of mixed-integer solution methods. 

Binary variables can be incorporated into continuous optimization solvers by enforcing equality constraints such as $x(x-1) = 0$, which forces $x$ to take a value in $\{0,1\}$, but such approaches are well known to perform poorly in practice. Alternatively, binary variables can be approximated using log-sum-exp functions~\cite[Chap.~3]{boyd2004convex}, sigmoid functions~\cite{iliev2017approximation}, or other similar techniques. These approximations, however, trade off numerical performance for accuracy and are not guaranteed to satisfy the desired logic exactly. In some situations, continuous relaxations of mixed-integer problems can yield feasible solutions without enforcing the integrality constraints~\cite{gunluk2010perspective}, but these relaxations are limited in scope and cannot be used for the general nonlinear optimal control problems we are interested in.

Temporal logic constraints provide another alternative to describe logic in optimization, enabling straightforward representations of time-dependent behavior~\cite{wolff2014optimization}. Nonetheless, standard techniques for solving problems with temporal logic constraints typically involve converting them into mixed-integer problems, which face the challenges discussed earlier, or rely on specialized satisfiability modulo theory (SMT) solvers, which struggle with nonlinear optimization problems \cite{belta2019formal}. Similarly, constraint programming can be applied to solve feasibility problems with complex constraints, but it does not scale well when dealing with systems that have large numbers of continuous variables.

The most relevant methods to our problem are presented in the recent works of \cite{malyuta2023fast} and \cite{cafieri2023continuous}, which propose different approaches for handling logical implications of the form of $g_1(x) \leq 0 \implies g_2(x) \leq 0$ in the constraints. In \cite{malyuta2023fast}, the authors integrate implication constraints into a sequential quadratic programming framework using a homotopy map, thus avoiding the need for binary variables. On the other hand, \cite{cafieri2023continuous} introduces a smooth piecewise-quadratic penalty function to enforce the implication constraints directly using continuous optimization solvers and demonstrates the approach's effectiveness on a feasibility problem.

\subsection{Contributions}

In this paper, we describe how to reformulate any logic-constrained optimization problem as a smooth optimization problem with the same solution. Our method converts the logic constraints to max-min constraints, then uses the often overlooked result of \cite{kirjner1998conversion} to replace them with an exact smooth substitute. We compare our proposed approach to several existing techniques for logic smoothing on two nonlinear control problems, and demonstrate how it achieves better convergence and faster numerical performance. The method described in this paper was previously utilized in our work on robust optimal control in \cite{wehbeh2024semi}, but is formalized and expanded upon here.

\section{Problem Description}

Consider a logic-constrained discrete-time optimal control problem between time steps $0$ and $N$ with state trajectory $x \coloneqq (x_0,x_1,x_2,\ldots,x_N) \in \mathbb{R}^n$ and control trajectory $u \coloneqq (u_0,u_1,u_2,\ldots,u_{N-1}) \in \mathbb{R}^m$. This problem can be formulated as

\begin{subequations}
\label{eq:logic_constrained_ocp}
\begin{equation}
\label{eq:logic_constrained_ocp_cost}
    \min_{x,u} \; J(x,u) 
\end{equation}
s.t.
\begin{align}
g(x,u) &\leq 0 \\
\label{eq:logic_constrained_ocp_eqs}
h(x,u) &= 0 \\
\label{eq:logic_constr_true}
L(x,u) \ &\text{is}\ \textbf{true}
\end{align}
\end{subequations}
where $J(\cdot,\cdot) : \mathbb{R}^n \times \mathbb{R}^m \rightarrow \mathbb{R}$ is the cost function to be minimized, $g(\cdot,\cdot) : \mathbb{R}^n \times \mathbb{R}^m \rightarrow \mathbb{R}^{n_g}$ is a set of inequality constraints that the solution to the optimization problem must satisfy, $h(\cdot,\cdot) : \mathbb{R}^n \times \mathbb{R}^m \rightarrow \mathbb{R}^{n_h}$ is a similar set of equality constraints, and $L(\cdot,\cdot) : \mathbb{R}^n \times \mathbb{R}^m \rightarrow \{\textbf{true},\textbf{false}\}$ is a logic program whose output is required to be \textbf{true}. The problem of~\eqref{eq:logic_constrained_ocp} only differs from a standard optimal control formulation by the addition of the logic program of \eqref{eq:logic_constr_true}. The formulation of \eqref{eq:logic_constrained_ocp_cost} to \eqref{eq:logic_constrained_ocp_eqs} is general enough to describe any optimal control problem, and we will soon show that \eqref{eq:logic_constr_true} can encode any logic requirements on $x$ and $u$. 

Here, the constraint $h(x,u)$ incorporates the traditional system dynamics, as well as any other equality constraints that are to be satisfied. If the discrete-time dynamics of the system are expressed as
    \begin{equation}
        x_{k+1} = f(x_k,u_k) \quad \forall k \in \{0,\ldots,N-1\}
    \end{equation}
    where $f(\cdot,\cdot) : \mathbb{R}^{n/(N+1)} \times \mathbb{R}^{m/N} \rightarrow \mathbb{R}^{n/(N+1)}$, then
    \begin{equation}
        h(x,u) = \begin{bmatrix}
            x_0 - \bar{x} \\
            x_1 - f(x_0,u_0) \\
            x_2 - f(x_1,u_1) \\
            \vdots \\
            x_N - f(x_{N-1},u_{N-1}) \\
            \bar{h}(x,u)
        \end{bmatrix}
    \end{equation}
where $\bar{x} \in \mathbb{R}^{n/(N+1)}$ represents the known initial conditions of the system and $\bar{h}(\cdot,\cdot) : \mathbb{R}^n \times \mathbb{R}^m \rightarrow \mathbb{R}^{\bar{n}_h}$ describes any other equality constraints applied to the system. 

Note that throughout this paper, we use the notation that is common to optimal control problems, where $x$ is the system's state and $u$ is the choice of control. However, the methods described here can be generalized to any optimization problem with decision variable $z = (x,u)$.

Next, we proceed to define the allowable forms for $L(\cdot,\cdot)$. Let $Q(\cdot,\cdot) : \mathbb{R}^n \times \mathbb{R}^m \rightarrow \{\textbf{true},\textbf{false}\}$ be an equality proposition that is only \textbf{true} when its associated function $q(x,u)$ takes a value of 0, where $q(\cdot,\cdot) : \mathbb{R}^n \times \mathbb{R}^m \rightarrow \mathbb{R}$. Similarly, let $P(\cdot,\cdot) : \mathbb{R}^n \times \mathbb{R}^m \rightarrow \{\textbf{true},\textbf{false}\}$ be an inequality proposition that is \textbf{true} when its associated function $p(\cdot,\cdot) : \mathbb{R}^n \times \mathbb{R}^m \rightarrow \mathbb{R}$ satisfies $p(x,u) \leq 0$, and \textbf{false} when $p(x,u) > 0$.

\begin{assum}
\label{assum:prob_form}
Let $\mathcal{L}$ be the set of all logical problems with $n_q \in \mathbb{N}_0$ equality propositions $Q_i$, $i \in \{1,\ldots,n_q\}$ and $n_p \in \mathbb{N}_0$ inequality propositions $P_j$, $j \in \{1,\ldots,n_p\}$. These propositions are linked by a finite number of logical operators NOT ($\neg$), AND ($\land$), and OR ($\lor$), such that $L$ outputs a single boolean truth value. Then, for the rest of this paper, we assume that $L \in \mathcal{L}$. 
\end{assum}

Allowing for the $\neg$, $\land$, and $\lor$ operators in the definition of $L$ is sufficient to define any relation between clauses, since this combination of operators is functionally complete~\cite[Thm~15D]{enderton2001mathematical}. This property, combined with equality and inequality propositions, allows $\mathcal{L}$ to describe any possible requirements on $x$ and $u$. Of course, this also means that $L$ can easily be structured to also include $g$ and $h$, but we choose to keep these constraints separate to differentiate them from constraints that cannot be expressed outside of $L$.

\section{Smooth Constraint Reformulations}

We now proceed to describe how any problem in $\mathcal{L}$ can be smoothed out to eliminate the need for constraint \eqref{eq:logic_constr_true} in the optimal control problem formulation.

\subsection{Logic Simplification}

\begin{lemma}
\label{lemma:no_neg_eq}
 Any problem within $\mathcal{L}$ can be rewritten using inequality propositions, $\land$, and $\lor$ operators exclusively.    
\end{lemma}
\begin{proof}
    In order to prove this result, we begin by showing that equality propositions can be replaced by an appropriate choice of inequality propositions. This can be done by introducing
    \begin{subequations}
    \begin{equation}
        P^+_{n_p +i} \land P^-_{n_p +i} \equiv Q_i \qquad \forall i \in \{1,\ldots,n_q\}
    \end{equation}
    with
    \begin{align}
        p_{n_p+i}^+(x,u) &\coloneqq q_i(x,u) \\
        p_{n_p+i}^-(x,u) &\coloneqq -q_i(x,u)
    \end{align}
    \end{subequations}
    or, alternatively, by introducing
    \begin{subequations}
    \begin{equation}
        P_{n_p +i} \equiv Q_i \qquad \forall i \in \{1,\ldots,n_q\}
    \end{equation}
    with
    \begin{equation}
        p_{n_p+i}(x,u) \coloneqq q_i^2(x,u).
    \end{equation}
    \end{subequations}
    The statement $P^+_{n_p +i} \land P^-_{n_p +i}$ is \textbf{false} if $q(x,u)$ is positive because of $P^+_{n_p +i}$ or if $q_i(x,u)$ is negative because of $P^-_{n_p +i}$. Therefore, $P^+_{n_p +i} \land P^-_{n_p +i}$ is only \textbf{true} when $q_i(x,u) = 0$. Similarly, $p_{n_p+i}(x,u)$ is positive for any non-zero values of $q_i(x,u)$, meaning $P_{n_p + i}$ is only \textbf{true} when $q_i(x,u) = 0$. The two approaches offer different numerical advantages and disadvantages, and can be applied interchangeably based on the user's preference.

    Following that, we push any $\neg$ operators to the level of the individual propositions using De Morgan's laws and simplify any double negations by the application of the rules
    \begin{align}
        \neg (P_1 \land P_2) & = \neg P_1 \lor \neg P_2 \\   
        \neg (P_1 \lor P_2) &= \neg P_1 \land \neg P_2 \\
        \neg \neg P &= P.
    \end{align}
    Consequently, we know that any problem in $\mathcal{L}$ can be written as a combination of $\land$ and $\lor$ operators, as well as $n_p^*$ inequality propositions and negated inequality propositions. We then fully eliminate the negations by noticing that
    \begin{equation}
    \begin{split}
        &\neg(p_i(x,u) \leq 0) \iff (p_i(x,u) > 0) \\ & \qquad \qquad\forall(x,u) \in \mathbb{R}^n \times \mathbb{R}^m,\, i \in \{1,\ldots,n_p^*\}
    \end{split}
    \end{equation}
    and that therefore,
    \begin{subequations}
    \label{eq:neg_ineq_equiv}
    \begin{equation}
        \bar{P_i} \equiv\neg P_i \qquad \forall i \in \{1,\ldots,n_p^*\}
    \end{equation}
    with
    \begin{equation}
        \bar{p}_i \coloneqq - p_i + \epsilon
    \end{equation}
    \end{subequations}
    where
    \begin{equation*}
        \epsilon \in \mathbb{R}, \qquad0 < \epsilon < \delta \qquad \forall \delta > 0
    \end{equation*}
    is used to convert the strict inequality into an inequality to allow for compatibility with standard optimization tools, which do not normally support strict inequalities. It then follows that any negated inequality proposition can be replaced by an equivalent proposition without the negation using \eqref{eq:neg_ineq_equiv}, thus completing the proof.
\end{proof}

In practice, setting $\bar{p}_i = - p_i$ is sufficient, since solvers cannot distinguish between strict and non-strict inequality constraints, and $\epsilon$ is only required to prove Lemma \ref{lemma:no_neg_eq}. 

\subsection{Max-Min Reformulation}

Next, we attempt to move away from the logical operators $\land$ and $\lor$, which cannot inherently be represented in continuous optimization solvers, and replace them with a formulation using the continuous (but not necessarily differentiable) $\max$ and $\min$ operators. In order to do so, we introduce the following equivalence.

\begin{lemma}
\label{lemma:constraint_equiv_minmax}
    For any problem $L \in \mathcal{L}$, and for any choice of $(x,u) \in \mathbb{R}^n \times \mathbb{R}^m$ the following statements are equivalent:
    \begin{itemize}
        \item The constraint of \eqref{eq:logic_constr_true} holds
        \item The constraint
        \begin{equation}
        \label{eq:logic_constr_cnf}
            \bigwedge\limits_{i=0}^{n_\land} \bigvee_{j=0}^{n_\lor(i)} P_{i,j} \ \text{is} \ \textbf{true}
        \end{equation}
        holds, where \eqref{eq:logic_constr_cnf} is the conjunctive normal form of $L$ under Lemma \ref{lemma:no_neg_eq}. This form has $n_\land + 1$ clauses joined through $\land$ operators, with each clause consisting of $n_\lor(i) + 1$ inequality propositions joined through $\lor$ operators, where $i \in \{0,\ldots,n_\land\}$ corresponds to the number of the clause.
        \item  The constraint
        \begin{equation}
        \label{eq:logic_constr_maxmin}
            \max_{i \in \{0,\ldots,n_\land\}} \ \min_{j \in \{0,\ldots,n_{\lor} (i)\}}\ p_{i,j}(x,u) \leq 0
        \end{equation}
        holds, where $p_{i,j}$ in \eqref{eq:logic_constr_maxmin} corresponds to $P_{i,j}$ in \eqref{eq:logic_constr_cnf} $\forall i \in \{0,\ldots,n_\land\}$, $\forall j \in\{0,\ldots,n_{\lor} (i)\}$.
    \end{itemize}
\end{lemma}
\begin{proof}
    The equivalence between \eqref{eq:logic_constr_true} and \eqref{eq:logic_constr_cnf} is relatively straightforward. Under Lemma \ref{lemma:no_neg_eq}, the logic problem~$L$ can be rewritten without any $\neg$ operators and without any equality propositions. Additionally, we know that any logical formula can be rewritten in conjunctive normal form (as a conjunction of disjunction, or an AND of ORs) such that this form is \textbf{true} if and only if the original problem also is \textbf{true} \cite[Chap.~1]{enderton2001mathematical}. Therefore, \eqref{eq:logic_constr_true} has a conjunctive normal form that does not include any $\neg$ operators or equality propositions, as in \eqref{eq:logic_constr_cnf}.

    Now, we prove the equivalence between \eqref{eq:logic_constr_cnf} and \eqref{eq:logic_constr_maxmin}. From \cite[Chap.~1]{wolsey1999integer}, we know that the logical AND operator on a group of inequalities of the form $\left[p_0(x,u) \leq 0 \right] \land \left[p_1(x,u) \leq 0 \right] \land \cdots \land \left[p_{n_\land}(x,u) \leq 0 \right]$ is true if and only if the largest $p_i(x,u)$, $i \in {0,\ldots,n_\land}$ satisfies $p_i(x,u) \leq 0$. Similarly, we know that $\left[p_0(x,u) \leq 0 \right] \lor \left[p_1(x,u) \leq 0 \right] \lor \cdots \lor \left[p_{n_\lor}(x,u) \leq 0 \right]$ is true if and only if the smallest $p_i(x,u)$, $i ={0,\ldots,n_\lor}$ satisfies $p_i(x,u) \leq 0$. Therefore, in \eqref{eq:logic_constr_cnf}, each logical OR over the propositions is equivalent to a $\min$, and the logical AND over the clauses is equivalent to a $\max$, proving that \eqref{eq:logic_constr_cnf} and \eqref{eq:logic_constr_maxmin} only hold if the other is true.
\end{proof}

Note that the total number of different propositions in the definitions of \eqref{eq:logic_constr_cnf} and \eqref{eq:logic_constr_maxmin} is equal to $n_p^*$, which is itself bounded between $n_p + n_q$ and $n_p + 2n_q$ depending on the approach used in eliminating the equality constraints. The conversion to conjunctive normal form does not involve introducing any new propositions and may only reshuffle or duplicate existing propositions. Therefore, the notation~$P_{i,j}$ simply represents a renumbering of the propositions to simplify the presentation. 

\subsection{Constraint Smoothing}

Before smoothing the constraint of \eqref{eq:logic_constr_maxmin}, we introduce the following assumption.

\begin{assum}
    \label{assum:smoothing}
    The functions $p_{i,j}(\cdot,\cdot)$ are twice differentiable $\forall i \in \{0,\ldots,n_\land\}$, $\forall j \in \{0,\ldots,n_\lor(i)\}$. Additionally, we require that their gradients $\nabla p_{i,j}(\cdot,\cdot)$ satisfy Assumption~3.9 in \cite{kirjner1998conversion}. This assumption essentially ensures that the gradients corresponding to each $\min$ operation are linearly independent and that the constraint and cost gradients satisfy a non-degeneracy condition.  
\end{assum}

Notice that the requirement for twice-differentiability in Assumption \ref{assum:smoothing} is less restrictive than may initially be thought, as most non-differentiable propositions can be split into multiple differentiable ones using logic when designing $L$. 

We now present the core result for this paper, which allows for the smoothing of any logic problem satisfying Assumptions \ref{assum:prob_form} and \ref{assum:smoothing}.

\begin{thrm}
    \label{thrm:smoothing_result}
    Any constraint of the form of \eqref{eq:logic_constr_true} for which $L \in \mathcal{L}$ satisfies Assumptions \ref{assum:prob_form} and \ref{assum:smoothing} can be replaced by
    \begin{equation}
    \label{eq:logic_constr_smooth}
        \sum_{j= 0}^{n_\lor(i)} \lambda_{i,j} p_{i,j}(x,u) \leq 0 \qquad  \forall i \in \{0,\ldots,n_\land\}
    \end{equation}
    without affecting the solution set, where $\lambda$ is a collection of auxiliary smoothing variables satisfying $(\lambda_{i,0},\lambda_{i,1},\ldots,\lambda_{i,n_\lor(i)}) \in \Lambda_{n_\lor(i)+1}$ $\forall i \in \{0,\ldots,n_\land\}$, $p_{i,j}(x,u)$, $n_\land$, and $n_\lor(\cdot)$ are defined as in \eqref{eq:logic_constr_maxmin}, and $\Lambda_k$ is the $k-$simplex set such that
    \begin{equation}
        \Lambda_k \coloneqq \left\{\lambda \in \mathbb{R}^k\left| \sum_{i=1}^k \lambda_i = 1,\, \lambda_i \geq 0 \right.\right\}.
        \vspace{5pt}
    \end{equation}
\end{thrm}

\begin{proof}
    From Lemma \ref{lemma:constraint_equiv_minmax}, we know that \eqref{eq:logic_constr_true} can be replaced by the constraint of \eqref{eq:logic_constr_maxmin}. Then under Assumption \ref{assum:smoothing}, we can apply the smoothing approach of \cite{kirjner1998conversion} to separate out the finite $\max$ into $n_\land +1$ different constraints and smooth each $\min$ to obtain the form of \eqref{eq:logic_constr_smooth}.
\end{proof}

The constraint of \eqref{eq:logic_constr_smooth} is differentiable up to the order of differentiability of each function $p_{i,j}(x,u)$, and is therefore at least twice differentiable. 

\subsection{Smooth Optimal Control Problem Formulation}

Under Theorem \ref{thrm:smoothing_result}, problem \eqref{eq:logic_constrained_ocp} can be reformulated as a standard optimal control problem with identical solution set
\begin{subequations}
\label{eq:smooth_ocp}
\begin{equation}
    \min_{x,u,\lambda} \; J(x,u) 
\end{equation}
s.t.
\begin{align}
g^*(x,u,\lambda) &\leq 0 \\
h^*(x,u,\lambda) &= 0
\end{align}
\end{subequations}
where $g^*(\cdot,\cdot,\cdot) : \mathbb{R}^n \times R^m \times \prod_{i = 0}^{n_\land} \Lambda_{n_\lor(i)} \rightarrow \mathbb{R}^{n_g^*}$ combines the constraints of $g$ with smoothed logic constraints and the positivity constraints for the $\lambda$ variables, $h^*(\cdot,\cdot,\cdot) : \mathbb{R}^n \times R^m \times \prod_{i = 0}^{n_\land} \Lambda_{n_\lor(i)} \rightarrow \mathbb{R}^{n_h^*}$ combines the constraints of $h$ with the unity constraints in the definition of the $\Lambda$ sets, $n_g^* =n_g + n_\land + 1 +\sum_{i = 0}^{n_\land} (n_\lor(i)+1)$, and $n_h^* =n_h + n_\land + 1$.

This problem formulation is easily differentiable and can be solved using standard continuous optimization algorithms as we will show in Section \ref{sec:results}.

It is worth noting that the conversion to conjunctive normal form is not always necessary during constraint smoothing. In some situations, it can be numerically advantageous to convert the logical operators to $\min$ and $\max$ operators directly, as is done in the examples of Section \ref{sec:results}.

\section{Numerical Simulations}
\label{sec:results}
We now illustrate the performance of our smoothing approach through two different control examples on a planar quadrotor problem.

\subsection{System Description}

\begin{figure}[t]
    \centering
    \includegraphics[width=0.5\columnwidth]{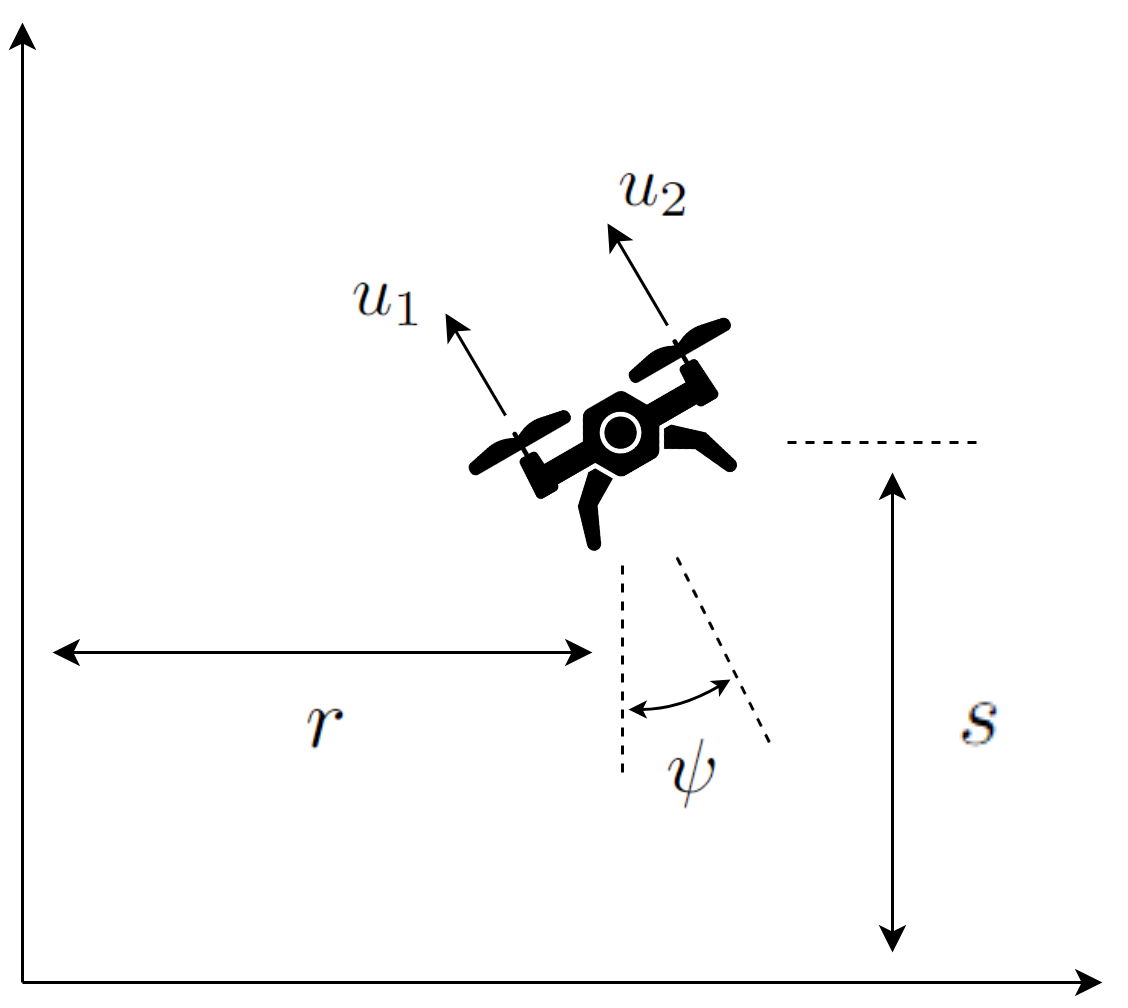}
    \vspace{-5pt}
    \caption{Illustration of the quadrotor's horizontal position ($r$), altitude ($s$), and tilt angle ($\psi$) taken from \cite{wehbeh2024robust}.}
    \label{fig:quad_example}
\end{figure}

Consider the planar quadrotor vehicle with states $[r,\dot{r},s,\dot{s},\psi,\dot{\psi}]$, where $r$ is the quadrotor's horizontal position, $s$ is the quadrotor's height, and $\psi$ is the quadrotor's tilt angle, as illustrated in Figure \ref{fig:quad_example}. The continuous-time dynamics for the system are 
\begin{equation}
    \begin{bmatrix}
        \ddot{r}(t) \\ \ddot{s}(t) \\ \ddot{\psi}(t)
    \end{bmatrix}
    =
    \begin{bmatrix}
        \sin(\psi(t)) \left(u_1(t) + u_2(t) \right) /\gamma \\
        \cos(\psi(t)) \left(u_1(t) + u_2(t) \right) /\gamma - g_{\earth} \\
        \ell (u_1(t) -u_2(t)) / I
    \end{bmatrix}
\end{equation}
where $u_1(t)$ and $u_2(t)$ are the thrusts at time $t$, $\gamma = 0.15$ is the vehicle's mass, $I = 0.00125$ is the moment of inertia, $\ell = 0.1$ is the motor moment arm, and $g_{\earth} = 9.81$ is the gravity acting on the system. Using a sampling time $T_s = 0.25$, the position dynamics are discretized via the trapezoidal rule and the velocity dynamics via explicit Euler, such that
\begin{equation}
\label{eq:ex_state_dyn}
    \begin{bmatrix}
        {x}^1_{k+1} \\
        {x}^2_{k+1} \\
        {x}^3_{k+1} \\
        {x}^4_{k+1} \\
        {x}^5_{k+1} \\
        {x}^6_{k+1} 
    \end{bmatrix}
    =
    \begin{bmatrix}
        {x}^1_{k} \\
        {x}^2_{k} \\
        {x}^3_{k} \\
        {x}^4_{k} \\
        {x}^5_{k} \\
        {x}^6_{k} 
    \end{bmatrix} +
    T_s
    \begin{bmatrix}
    \left(x_{2,k} + x_{2,k+1} \right)/2 \\[3pt]
    \sin(x_k^5) \left(v^1_k + v_k^2 \right) /\gamma \\[3pt]
    \left(x_{4,k} + x_{4,k+1} \right)/2 \\[3pt]
    \cos(x_k^5) \left(v^1_k + v_k^2 \right) /\gamma - g_{\earth} \\[3pt]
    \left(x_{6,k} + x_{6,k+1} \right)/2 \\[3pt]
    \ell (v^1_k -v^2_k) / I
    \end{bmatrix}
\end{equation}
where $x^1_k$ through $x^6_k$ are the discretized states corresponding to the 6 continuous states at time step $k$, $v_k^1$ and $v_k^2$ are the discretized control inputs, and $k = 0$ corresponds to $t = 0$. For both of the examples of Section \ref{sec:ocp_description}, we consider a prediction horizon of $N = 10$. The quadrotor is initialized from the state $x_0 = [0,0,0,0,0,0]$ and is subject to the input constraints
\begin{equation}
    -2 \leq v_k^i \leq 2 \qquad \forall i \in \{1,2\}.
\end{equation}

\subsection{Logic-Constrained Optimal Control Problems}
\label{sec:ocp_description}

\begin{figure*}[t]
  \centering
  \begin{tabular}{ c @{\hspace{20pt}} c @{\hspace{20pt}} c}
    \includegraphics[width=.45\columnwidth,trim = {0 10pt 0 20pt}]{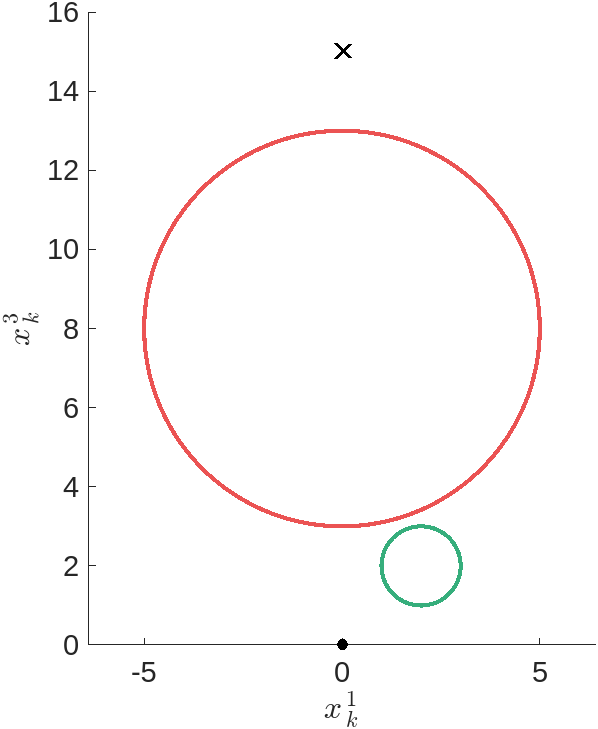} \hspace{20pt} &
    \includegraphics[width=.45\columnwidth,trim = {0 10pt 0 20pt}]{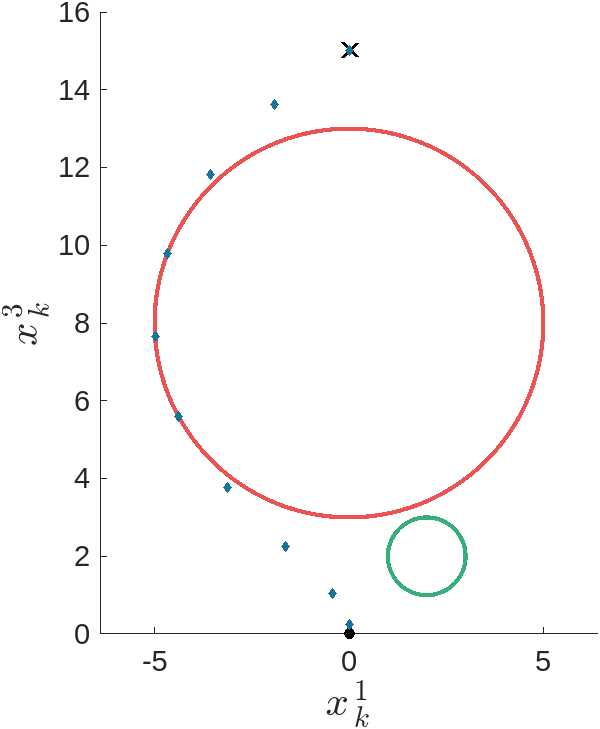} \hspace{20pt} & 
    \includegraphics[width=.45\columnwidth,trim = {0 10pt 0 20pt}]{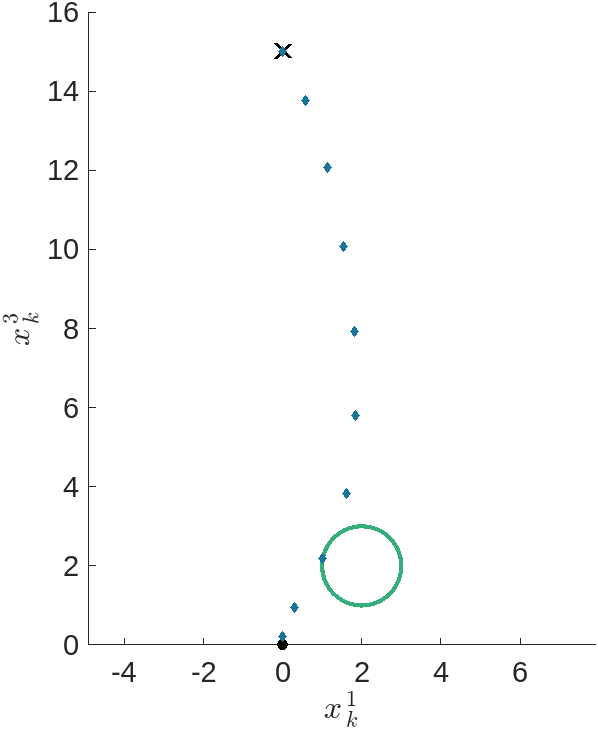} \\[5pt]
    \parbox[b]{.54 \columnwidth}{\footnotesize (a) \ Going through the green circle allows passing through the red obstacle.} &
      \parbox[b]{.54 \columnwidth}{\footnotesize (b) \ Sample sub-optimal trajectory. \phantom{filler} \phantom{filler} \phantom{filler} \phantom{filler}} &
      \hspace{15pt} \parbox[b]{.54 \columnwidth}{\footnotesize (c) \ Optimal solution trajectory. \phantom{filler} \phantom{filler} \phantom{filler} \phantom{filler}}
  \end{tabular}

  \smallskip

  \caption{Obstacle (red), logic trigger (green), target (black), and solution trajectory (blue) for Problem 1.}
  \label{fig:prob_1}
\end{figure*}

\begin{figure*}[t]
  \centering
  \begin{tabular}{ c @{\hspace{20pt}} c @{\hspace{20pt}} c}
    \includegraphics[width=.45\columnwidth,trim = {0 10pt 0 20pt}]{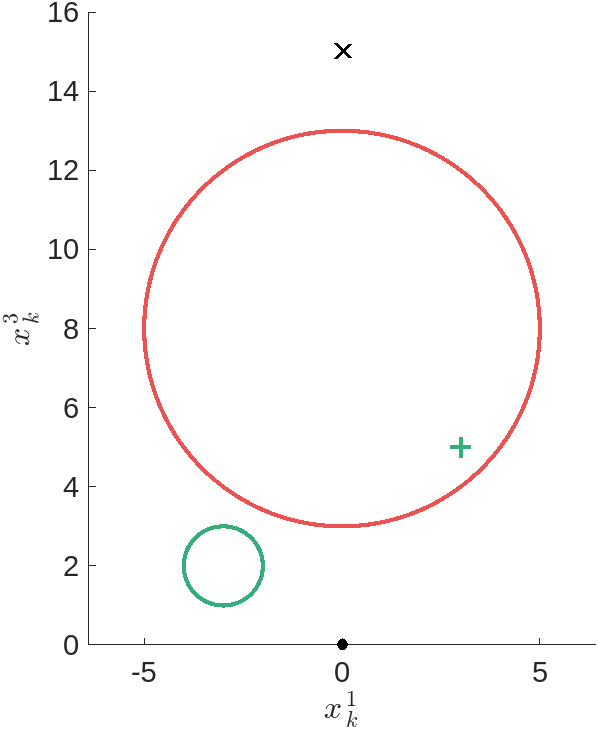} \hspace{20pt} &
    \includegraphics[width=.45\columnwidth,trim = {0 10pt 0 20pt}]{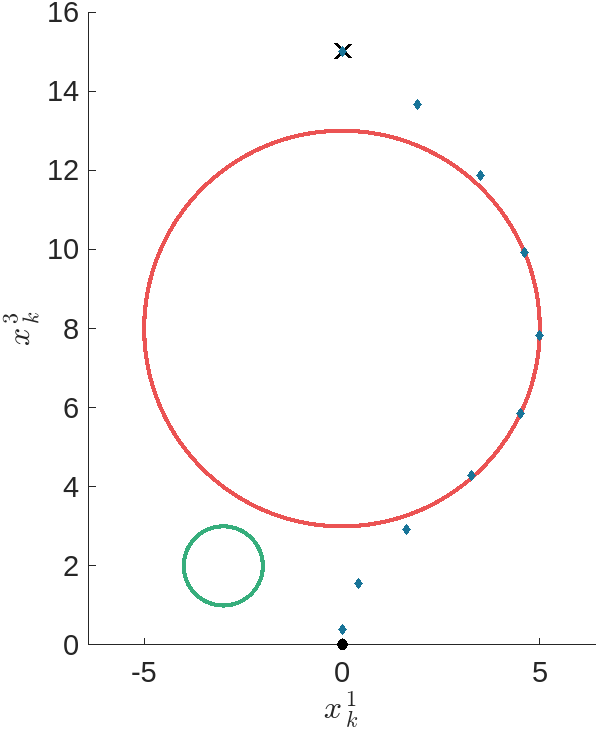} \hspace{20pt} & 
    \includegraphics[width=.45\columnwidth,trim = {0 10pt 0 20pt}]{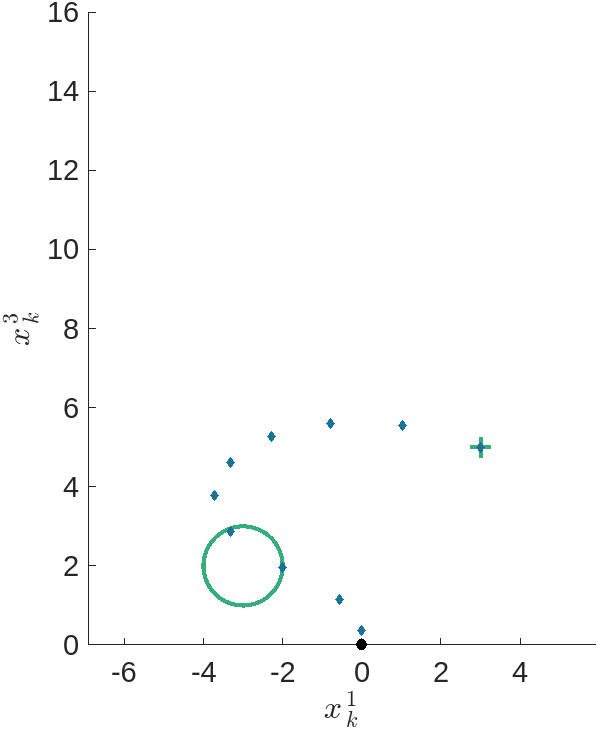} \\[5pt]
    \parbox[b]{.58 \columnwidth}{\footnotesize (a) \ Going through the green circle removes the red obstacle and switches to the + target.} &
      \parbox[b]{.54 \columnwidth}{\footnotesize (b) \ Sample sub-optimal trajectory. \phantom{filler} \phantom{filler} \phantom{filler} \phantom{filler} \phantom{filler}} &
      \hspace{15pt} \parbox[b]{.54 \columnwidth}{\footnotesize (c) \ Optimal solution trajectory. \phantom{filler} \phantom{filler} \phantom{filler} \phantom{filler}  \phantom{filler}}
  \end{tabular}

  \smallskip

  \caption{Obstacle (red), logic trigger (green), targets (black x and green +), and solution trajectory (blue) for Problem 2.}
  \label{fig:prob_2}
\end{figure*}

\subsubsection{Problem 1}
The first problem we consider minimizes the cost function
\begin{equation}
    J(v) = \sum_{i=0}^{N-1} (v_i^1)^2 + (v_i^2)^2
    \label{eq:prob_1_cost}
\end{equation}
subject to the terminal constraints $x_N^1 = 0$ and $x_N^3 = 15$.
The problem also has to satisfy
\begin{equation}
\begin{split}
    & \textbf{IF}\ \textbf{NOT} \left\{ (x_i^1 - 2)^2 + (x_i^3 - 1)^2 - 1 \leq 0, \, \forall i \in \{2,3\}  \right\} \\ & \textbf{THEN} \ \textbf{NOT} \left\{ (x_i^1)^2 + (x_i^3 - 8)^2 - 25 \leq 0, \, \forall i \in \{5,\ldots,9\}  \right\} 
\end{split}
\raisetag{28pt}
\label{eq:problem_1_logic}
\end{equation}
which can be interpreted as requiring the quadrotor to avoid the red circle in Figure \ref{fig:prob_1}\,(a) unless it passes through the green circle at time steps 2 or 3. Since $\textbf{IF}\ A \ \textbf{THEN} \ B$ is logically equivalent to $\neg A \lor B$, we can rewrite \eqref{eq:problem_1_logic} using $\neg$, $\land$, and $\lor$ operators in the form of $\mathcal{L}$. By applying the procedure of Theorem \ref{thrm:smoothing_result}, the constraint then becomes
\begin{equation}
\begin{split}
    &\lambda_{1} \left( (x_2^1 - 2)^2 + (x_2^3 - 1)^2 - 1  \right) \\
    &\qquad + \lambda_{2} \left( (x_3^1 - 2)^2 + (x_3^3 - 1)^2 - 1  \right) \quad \ \  \forall i \in \{5,\ldots,9\}\\
    & \qquad- \lambda_{3} \left( (x_i^1)^2 + (x_i^3 - 8)^2 - 25  \right ) \leq 0
\end{split} 
\raisetag{0.75\baselineskip}
\end{equation}
replacing the logic problem by a set of 5 smooth constraints. 

\subsubsection{Problem 2} 

The second problem we consider shares the same cost of \eqref{eq:prob_1_cost} as Problem 1, but replaces 
\eqref{eq:problem_1_logic} by the new logic constraint
\begin{equation}
\label{eq:prob_2_logic}
    \begin{split}
    &\textbf{IF} \left\{ (x_3^1 + 3)^2 + (x_3^3 - 2)^2 - 1 \leq 0  \right\} \\
    & \qquad \qquad\textbf{THEN} \ \{ x_N^1 = 3 \ \textbf{AND} \ x_N^3 = 5 \} \\
    &\textbf{ELSE} \left\{ \textbf{NOT} \left\{ (x_i^1)^2 + (x_i^3 - 8)^2 - 25 \leq 0, \, \forall i \in \{5,\ldots,9\} \right\} \hspace{-3pt} \right. \\
    & \left. \qquad \qquad \textbf{AND} \ \{ x_N^1 = 0 \ \textbf{AND} \ x_N^3 = 15 \} \right\}
    \end{split}
    \raisetag{44pt}
\end{equation}
which, as seen in Figure \ref{fig:prob_2}\,(a), requires the quadrotor to reach the black x target while avoiding the red obstacle, unless it is inside the green circle at time step 3, at which point it must instead reach the green + target. This constraint is smoothed using the same procedure as for Problem 1. 

\subsection{Numerical Results}

Each of the problems is solved in Julia using the JuMP package \cite{Lubin2023} and the Ipopt optimizer \cite{wachter2006implementation} on a laptop with an 11th Gen Intel\textsuperscript{\tiny\textregistered} Core\textsuperscript{\texttrademark} i7-11370H CPU at 3.30 GHz and 16GB of RAM. For Problem 1, we compare our solution method to an approach that enforces the binary variables through a complementarity condition, and to a Big M method that represents the condition $\{a \leq 0\} \lor \{b \leq 0\} \lor \{c \leq 0\}$ as $\{a \leq \mu_1M\} \land \{b \leq \mu_2 M\} \land \{c \leq \mu_3 M\}$ with continuous variables $\mu$ such that $\mu_1 \mu_2 \mu_3 = 0$. We do not implement these methods for Problem 2 because using them to express the logic of \eqref{eq:prob_2_logic} leads to a large number of binaries and poor numerical behavior. Each approach is evaluated on 1000 different runs with randomized initial guesses. 

\subsubsection{Problem 1} The optimal solution to Problem 1 requires flying through the logic trigger, as seen in Figure \ref{fig:prob_2}\,(c). The results of Table \ref{tab:prob_1_perf} demonstrate that our method outperforms the other approaches, returning the optimal solution 81.3\% of the time and failing to produce a feasible trajectory across only 4.4\% of runs. The smoothed constraints also display the fastest computation times, even when the slower infeasible runs are excluded. We note that sub-optimal solutions are still observed despite the exactness of the smoothing because of the non-convex nature of the optimization problem.

\begin{table}[h]
\centering
\vspace{5pt}
\caption{Performance Metrics on Problem 1 }
\vspace{-7pt}
\label{tab:prob_1_perf}
\begin{tabular}{@{}lcccc@{}}
\toprule
                & Opt. \# & Sub-Opt.  \# & Inf. \# & Avg. Cost \\ \midrule
Smoothed        & 813         & 143              & 44                 & 22.24     \\
Complementarity & 325         & 518              & 157                & 28.67     \\
Big M           & 660         & 274              & 66                 & 23.73     \\ \midrule
                & Avg. Time & \multicolumn{2}{c}{Avg. Time (Feas.)} & Max Time \\ \midrule
Smoothed        & 22.4\,ms     & \multicolumn{2}{c}{19.4\,ms}              & 466.8\,ms   \\
Complementarity & 76.1\,ms      & \multicolumn{2}{c}{78.2\,ms}              & 419.7\,ms  \\
Big M           & 56.7\,ms      & \multicolumn{2}{c}{37.9\,ms}              & 1782.4\,ms   \\ \bottomrule
\end{tabular}
\end{table}

\subsubsection{Problem 2} The optimal solution to Problem 2 similarly requires flying through the green logic trigger, as seen in Figure \ref{fig:prob_2}\,(c). The smoothed constrained approach returns feasible trajectories 92.3\% of the time, and optimal solutions for 71.7\% of runs. A full overview of the method's performance can be seen in Table \ref{tab:prob_2_perf}. 

\begin{table}[h]
\centering
\vspace{5pt}
\caption{Performance Metrics on Problem 2}
\vspace{-7pt}
\label{tab:prob_2_perf}
\begin{tabular}{@{}lcccc@{}}
\toprule
                & Opt. \# & Sub-Opt.  \# & Inf. \# & Avg. Cost \\ \midrule
Smoothed        & 717         & 196              & 87                 & 25.62    \\ \midrule
                & Avg. Time & \multicolumn{2}{c}{Avg. Time (Feas.)} & Max Time \\ \midrule
Smoothed        & 122.2 ms     & \multicolumn{2}{c}{106.7 ms}              & 1840.1 ms  \\ \bottomrule
\end{tabular}
\end{table}

\section{Conclusion}

In this paper, we presented a method for reformulating arbitrary logic constraints as continuously differentiable expressions without introducing any conservativeness. Our approach enables the formulation of complex logic within continuous optimization solvers, eliminating the need for binary variables. We compared our method to existing techniques on two quadrotor control problems involving conditional equality and inequality constraints, and observed significant improvements in both performance (81.3\% vs 66\% optimal solution rate) and computation time (22.4\,ms vs. 56.7\,ms average). Finding the representation of the logic problem that leads to the best performance after smoothing remains an open question, and may be explored further in future work. 

\section*{ACKNOWLEDGMENTS}

This work was funded by the Natural Sciences and Engineering Research Council of Canada through a PGS D grant. 




\bibliographystyle{IEEEtran}
\bibliography{IEEEabrv,references.bib}

\end{document}